\documentclass[showpacs,preprintnumbers,epsf,axodraw,
superscriptaddress,amsmath]{revtex4}

\usepackage{epsfig}

\newcommand\e{\epsilon}

\newcommand\p{\pi}

\newcommand{\be}{\begin{equation}}
\newcommand{\ee}{\end{equation}}
\newcommand{\bea}{\begin{eqnarray}}
\newcommand{\eea}{\end{eqnarray}}
\newcommand{\ba}[1]{\begin{array}{#1}}
\newcommand{\ea}{\end{array}}

\newcommand{\eq}[1]{Eq.\,(\ref{#1})}
\newcommand{\eqs}[2]{Eqs.\,(\ref{#1})-(\ref{#2})}
\newcommand{\eqq}[2]{Eqs.\,(\ref{#1}) and (\ref{#2})}

\begin{document}

\title{Gluon self-energy effect on CFL gap equation}

\author{H.\ Malekzadeh}
\email{malekzadeh@uni-muenster.de}
\affiliation{Institut f\"ur Theoretische Physik,
Westf\"alische Wilhelms-Universit\"at M\"unster, Wilhelm-Klemm-Strasse\ 9,
48149 M\"unster, Germany}

\date{\today}

\begin{abstract}
I investigate the effects of the gluon self-energy in the color-flavor-locked phase of color superconductivity on the value of the gap. The light plasmon modes, which appear for energies smaller than twice the gap energy, provide the largest contribution. They modify the subsubleading term in the solution of the gap equation.

\end{abstract}

\maketitle
\section{Introduction}

At very high densities, $\mu\gg\Lambda_{QCD}$, single-gluon exchange is the dominant interaction between quarks. This interaction is attractive in the color-antitriplet channel. Therefore, sufficiently cold and dense quark matter is a color superconductor \cite{1}. At such densities, asymptotic freedom \cite{2} implies that the strong coupling constant, $g$, is much smaller than unity, $g(\mu)\ll 1$. Hence, in this limit, the calculations of the color-superconducting gap parameter $\phi$ can be controlled.

In order to compute the gap parameter one has to solve the gap equation, which is in general a function of the fermionic quasiparticles four-momentum $K_\mu\equiv(k_0,{\bf k})$. For an ordinary superconductor, in the weak coupling limit, $p_D\sim g\mu\ll\mu$, where $p_D$ is the Debye momentum, the the solution of the gap equation is
\bea\label{bcsgap}
\phi=b_{BCS}\,\mu\,\exp\left(-\frac{c_{BCS}}{g^2}\right)\,.
\eea
Here, $b_{BCS}\equiv2\omega_D/\mu$ is a dimensionless constant and $\omega_D$ is the Debye frequency. To derive \eq{bcsgap} it is assumed that the attractive phonon exchange interaction between electrons is local \cite{4,5}. The locality for the exchanged particles, however, is not a crucial condition. One can generalize the solution to nonlocal interactions of finite range, for instance, to the massive scalar boson exchange \cite{6}. If we assume that fermions are massless and the boson mass is generated by in-medium effects, $M_B\simeq g\mu$, the energy scales in the weak coupling limit are $\phi\ll M_b\ll\mu$. The solution of the gap equation is then
\bea\label{bcsgapp}
\phi\simeq b_B\,\mu\exp\left(-\frac{c_B}{g^2{\rm ln}(2\mu/M_B)}\right)\,.
\eea
The main difference of this solution with \eq{bcsgap} is that the gap becomes larger, i.e. $g^2$ is replaced by $g^2{\rm ln}(2\mu/M_B)\sim g^2{\rm ln}(1/g)\gg g^2$.

In QCD gluon exchange is a nonlocal interaction. In the vacuum gluons are massless; hence, the interaction is of infinite range. In the dense medium, however, the electric and the nonstatic magnetic gluon exchanges are screened. In the latter case, the behavior of the gluons is very similar to the scalar bosons of the previous example with a mass
\bea\label{gmass}
m_g^2=N_f\frac{g^2\mu^2}{6\pi^2}+\left(N_c+\frac{N_f}{2}\right)\frac{g^2 T^2}{9},
\eea
where $N_f$ and $N_c$ are the number of massless quark flavors and quark colors participating in the pairing, respectively. $T$ is the temperature. Notice that, the exchange of the almost static magnetic gluons is still unscreened and thus has infinite range \cite{3}. In the weak coupling limit, all unscreened gluons contribute to leading order to the gap equation, whereas all screened gluons contribute to subleading order. All other contributions construct the subsubleading term \cite{rischke094}.

In comparison to the BCS theory, where the maximum phonon momentum is the Debye momentum $p_D$, in QCD there is no restriction on the magnitude of the gluon momentum in the gap equation. Nevertheless, in the weak coupling limit, the gap function is strongly peaked at the Fermi surface $q={\rm k}_F$, and this indirectly again restricts the gluon momentum to a narrow range around the Fermi surface, ${\rm k}_F-\delta\le q\le {\rm k}_F+\delta$ \cite{3,7,8}. Under this condition, the solution of the gap equation is
\bea\label{qcdgap}
\phi=b\,\mu\,\exp\left(-\frac{c}{g}\right)\left[1+O(g)\right]\,,
\eea
where $b=256\pi^4 (2/g^2 N_f)^{5/2} b'_0$. The dependence of $b$ on $g$ arises from the gluon mass, \eq{gmass}. At the present approximation $b'_0=1$ \cite{3,8,12,13}. Including a finite, $\mu$-dependent contribution to the quark wave function renormalization, Brown, Liu, and Ren in Ref.\,\cite{14} showed that
\bea
b'_0=\exp\left(-\frac{\pi^2+4}{8}\right)b''_0\simeq 0.176\, b''_0\,.
\eea
Using the standard running coupling constant in the vacuum, one can find $b''_0=\exp[33(\pi^2-4)]\simeq 20$ \cite{15}. We see that in comparison to \eqq{bcsgap}{bcsgapp} the power of the coupling constant $g$ in the exponent of \eq{qcdgap} is reduced \cite{9}. In the hard-dense-loop (HDL) limit, the constant $c$ in \eq{qcdgap} is calculated by Son \cite{10,11}:
\bea
c=\frac{3\pi^2}{\sqrt{2}}\,.
\eea
In the other works \cite{wr}, the effect of the quark self-energy is taken into account. This gives rise to a more accurate value for $b'_0$. Later on, the solution of the gap equation for different phases of color superconductivity were presented in several papers \cite{schmitt}. Recently, it is also shown that the subleading order contributions to the gap from the vertex corrections are absent for the gapped excitations \cite{wang}.

In this paper, at asymptotically large densities $\phi\ll m_g\ll\mu$, I investigate the effect of the gluon self-energy on the solution of the gap in the color-flavor-locked (CFL) phase of color superconductor. It is believed that this effect is of leading or subleading order \cite{rischke3063,007}. These calculations are for the complete range of the gluon energy and momentum $p_0,p\ll\mu$. I consider small temperatures $T\sim\phi\ll\mu$ where the dominant contribution to the one-loop gluon self-energy comes from the quark loops, which are $\sim g^2\mu^2$. At this limit, the contribution of the gluon and the ghost loops are suppressed, because they are proportional to $g^2 T^2$ \cite{007}.

The gluon self-energies in the CFL phase are explicitly calculated in Refs.\,\cite{rischke3063,mal1}. It is known that there is a light plasmon mode in the CFL phase. Therefore, the other aim of this paper is to know the effect of these unique modes on the value of the CFL gap. Similar calculations for the 2SC phase are given in Ref.\,\cite{rischke094}, except, in those calculations the light plasmon mode is not studied.


This paper is organized as follows. In Sec. \ref{gapequation}, I solve the gap equation in the color-flavor-locked phase. For that, I need to know the full gluon propagators. In Sec. \ref{gluonpropagators}, it is shown how one can write these propagators in terms of their spectral densities. The discussions about the different limits of the gluon spectral densities are presented in \ref{spectraldensities}. The form of the gap equation requires one to have knowledge about the dispersion relation of the gluons. This is explained in \ref{dispersionrelations}. Collecting all these materials, we become ready to estimate the effect of the gluon self-energy on the solution of the gap. Section \ref{qualitativeresults} is divided to three subsections. Subsections \ref{estimate00} and \ref{estimatet} are devoted to estimate the effect of the gluon self-energy in color-superconducting phase in comparison to that in the HDL limit. In \ref{sphi}, I achieve my goal and find the effect of the largest contribution from the gluon self-energy on the gap function. In the end, I conclude.

\section{Gap equation}\label{gapequation}

The gap equation for the color-superconducting condensate of the massless fermions is derived in \cite{6}. At nonzero temperature this equation reads
\bea\label{gapeq}
\Phi^+=g^2\frac{T}{V}\sum_{Q}\bar{\Gamma}^\mu_a\Delta^{ab}_{\mu\nu}(K-Q)G_0^-(Q)\Phi^+(Q)G^+(Q)\Gamma^\nu_b\,,
\eea
where a summation over Lorentz indices $\mu, \nu$ as well as adjoint color indices $a,b=1,\dots,8$ is implied, and in the finite-volume limit we have
\bea
\frac{T}{V}\sum_Q\equiv\sum_n\int\frac{d^3{\bf q}}{2\pi^3}\,.
\eea
Here $n$ labels the fermionic Matsubara frequencies
\bea
\omega_n\equiv(2n+1)\pi T\equiv i q_0\,.
\eea
The full gluon propagator is related to the gluon self-energy $\Pi^{ab}_{\mu\nu}$ via
\bea
[\Delta^{-1}]^{ab}_{\mu\nu}=[\Delta^{-1}_0]^{ab}_{\mu\nu}-\Pi^{ab}_{\mu\nu}\,,
\eea
where $[\Delta^{-1}_0]^{ab}_{\mu\nu}$ is the bare gluon propagator. The propagator for the free, massless particles (upper sign) or charge-conjugate particles (lower sign),
\bea
[G_0^\pm(Q)]^{-1}\equiv \gamma\cdot Q\pm\gamma_0\,\mu\,,
\eea
constitutes the propagator for the quasiparticles (upper sign) or the charge-conjugate quasiparticles (lower sign) as follows
\bea
[G^\pm(Q)]^{-1}\equiv[G_0^\pm (Q)]^{-1}-\Sigma^\pm(Q)\,.
\eea
The quark self-energy arises from the interaction with the condensate
\bea
\Sigma^\pm(Q)\equiv\Phi^\mp G_0^\mp \Phi^\pm\,.
\eea
The charge-conjugate condensate $\Phi^-$ is related to the condensate $\Phi^+$ via
\bea
\Phi^-\equiv\gamma_0 (\Phi^+)^\dagger\gamma_0\,,
\eea
and the vertices in \eq{gapeq} are defined as
\bea
\Gamma^\mu_a\equiv T_a\gamma^\mu\,\,\,,\,\,\, \bar{\Gamma}^\mu_a\equiv-\gamma^\mu T^T_a\,,
\eea
where $\gamma^\mu$ is the Dirac matrix and $T_a$ the Gell-Mann matrix.

Using the following projectors
\begin{subequations}\label{projectors}
\bea
\mathcal{C}^{(1)fg}_{\hspace{.4cm}ij}&\equiv&\frac{1}{3}\delta^f_i\delta^g_j\,,\\
\mathcal{C}^{(2)fg}_{\hspace{.4cm}ij}&\equiv&\frac{1}{2}\left(\delta^{fg}\delta_{ij}-\delta^f_j\delta^g_i\right)\,,\\
\mathcal{C}^{(3)fg}_{\hspace{.4cm}ij}&\equiv&\frac{1}{2}\left(\delta^{fg}\delta_{ij}+\delta^f_j\delta^g_i\right)-\frac{1}{3}\delta^f_i\delta^g_j\,,
\eea
\end{subequations}
from Ref.\cite{projectors}, the gap matrix \eq{gapeq} can be written as
\bea\label{gapss}
\Phi^\pm\equiv\sum_{n=1}^3\Phi^\pm_n
\eea
where
\begin{subequations}\label{gaps}
\bea
\Phi_1^\pm&\equiv& 2\big(\Phi_{\bar{3}}^\pm+2\Phi_6^\pm\big)\,,\\
\Phi_2^\pm&\equiv& \Phi_{\bar{3}}^\pm-\Phi_6^\pm\,,\\
\Phi_3^\pm&\equiv& -\Phi_2^\pm\,,
\eea
\end{subequations}
are gap matrices in the spinor space,
\bea\label{spinor}
\Phi_n^+&\equiv&\sum_{h=r,\ell}\sum_{e=\pm}\phi_{n,h}^e(K)\mathcal{P}_h\Lambda^e_{\bf k}\,,\\
\Phi_n^-&\equiv&\sum_{h=r,\ell}\sum_{e=\pm}[\phi_{n,h}^e(K)]^*\mathcal{P}_{-h}\Lambda^{-e}_{\bf k}\,.
\eea
The chirality projectors
\bea
\mathcal{P}_{r,\ell}\equiv\frac{1}{2}\big(1\pm\gamma_5\big)
\eea
are defined so that $-h=\ell$ when $h=r$ and $-h=r$ when $h=\ell$. In addition
\bea
\Lambda_{\bf k}^\pm\equiv\frac{1}{2}\big(1\pm\gamma_0\,\gamma\cdot\hat{{\bf k}}\big)
\eea
are energy projectors and $\phi_{n,h}^e(K)$ in \eq{spinor} is the gap function for the pairing of quarks $(e=+1)$ or antiquarks $(e =-1)$ with chirality $h$.

From \eq{gaps} one can write the gap matrix \eq{gapeq} in terms of the triplet $\Phi^\pm_{\bar{\bf 3}}$ and the sextet $\Phi^\pm_{6}$ gaps
\bea
[\Phi^\pm]^{fg}_{ij}\equiv\Phi^\pm_{\bar{\bf 3}}\big(\delta^f_i\delta^g_j-\delta^f_j\delta^g_i\big)+\Phi^\pm_{6}\big(\delta^f_i\delta^g_j+\delta^f_j\delta^g_i\big)\,.
\eea
For convenience we use the following notations:
\bea\label{phisym}
\Phi^\pm_{\bf 1}\equiv\Phi^\pm_1\,\,\,,\,\,\,\Phi^\pm_{\bf 8}\equiv\Phi^\pm_2\equiv-\Phi^\pm_3\,.
\eea
Hence, similar to \eq{gapss}, one can write the quasiparticle propagators in terms of the projectors introduced in \eq{projectors}
\bea
G^\pm(K)\equiv\sum_{n=1}^3 G^\pm_n\,,
\eea
where
\bea\label{pron}
G^\pm_n(K)\equiv\sum_{h=r,\ell}\sum_{e=\pm}\frac{\mathcal{P}_{\pm h}\Lambda^{\pm e}_{\bf k}}{k_0^2-[\e^e_{\bf k}(\phi_{n,h}^e)]^2}[G_0^\mp(K)]^{-1}\,.
\eea
Here, $\e^e_{\bf k}(\phi_{n,h}^e)$ is the quasiparticles energy
\bea
[\e^e_{\bf k}(\phi_{n,h}^e)]^2\equiv (\mu-ek)^2+|\phi_{n,h}^e|^2\,.
\eea
Using \eq{phisym} and \eq{pron} we have
\bea\label{propagn}
[G^\pm]^{fg}_{ij}\equiv [{\bf P_1}]^{fg}_{ij} G_{\bf 1}^\pm+[{\bf P_8}]^{fg}_{ij}G^\pm_{\bf 8}\,,
\eea
where
\begin{subequations}
\bea
\left[{\bf P_1}\right]^{fg}_{ij}&\equiv&\frac{1}{3}\delta^f_i\delta^g_j\,,\\
\left[{\bf P_8}\right]^{fg}_{ij}&\equiv&\delta^{fg}\delta_{ij}-\frac{1}{3}\delta^f_i\delta^g_j\,.
\eea
\end{subequations}
Combining \eqs{projectors}{gaps}, we can write a relation similar to that in \eq{propagn} for the gap matrix
\bea\label{gapproj}
\left[\Phi^\pm\right]^{fg}_{ij}\equiv[\mathcal{Q}_1]^{fg}_{ij}\Phi_{\bf 1}^\pm+[\mathcal{Q}_8]^{fg}_{ij}\Phi^\pm_{\bf 8}\,,
\eea
but with different projectors
\begin{subequations}
\bea
\left[\mathcal{Q}_1\right]^{fg}_{ij}&\equiv&\frac{1}{3}\delta^f_i\delta^g_j\,,\\
\left[\mathcal{Q}_8\right]^{fg}_{ij}&\equiv&\frac{1}{3}\delta^f_i\delta^g_j-\delta^f_j\delta^g_i\,.
\eea
\end{subequations}
We see that $[\mathcal{Q}_1]^{fg}_{ij}\equiv\left[{\bf P_1}\right]^{fg}_{ij}$. Furthermore, one can check that
\begin{subequations}\label{pproj}
\bea
\left[\mathcal{Q}_1\right]^{fg}_{ij}&\times&\left[{\bf P_1}\right]^{gh}_{jl}\equiv [\mathcal{Q}_1]^{fh}_{il}\,,\\
\left[\mathcal{Q}_1\right]^{fg}_{ij}&\times&\left[{\bf P_8}\right]^{gh}_{jl}\equiv 0\,,\\
\left[\mathcal{Q}_8\right]^{fg}_{ij}&\times&\left[{\bf P_1}\right]^{gh}_{jl}\equiv 0\,,\\
\left[\mathcal{Q}_8\right]^{fg}_{ij}&\times&\left[{\bf P_8}\right]^{gh}_{jl}\equiv [\mathcal{Q}_8]^{fh}_{il}\,,
\eea
\end{subequations}
together with
\begin{subequations}\label{ppproj}
\bea
\left[\mathcal{Q}_1\right]^{fg}_{ij}&\times&\left[\mathcal{Q}_1\right]^{gf}_{ji}\equiv 1\,,\\
\left[\mathcal{Q}_1\right]^{fg}_{ij}&\times&\left[\mathcal{Q}_8\right]^{gf}_{ji}\equiv 0\,,\\
\left[\mathcal{Q}_8\right]^{fg}_{ij}&\times&\left[\mathcal{Q}_8\right]^{gf}_{ji}\equiv 8\,.
\eea
\end{subequations}
Writing the gap equation, \eq{gapeq}, in terms of its color and flavor indices
\bea\label{gapeqcf}
\left[\Phi^+\right]^{fg}_{ij}(K)\hspace{-.2cm}&\,=\,\hspace{-.2cm}&g^2\frac{T}{V}\sum_{Q}\gamma^\mu(T^T_a)_{ic}\,\Delta^{ab}_{\mu\nu}(K-Q)\left[G_0^-(Q)\right]^{fl}_{cm}\left[\Phi^+(Q)\right]^{ln}_{mk}\left[G^+(Q)\right]^{ng}_{kd}(T_a)_{dj}\gamma^\nu
\eea
and using \eq{pproj}, the gap equation finds the form
\bea\label{gap2}
\left[\Phi^+\right]^{fg}_{ij}(K)\hspace{-.2cm}&\,=\,\hspace{-.2cm}&g^2\frac{T}{V}\sum_{Q}\gamma^\mu(T^T_a)_{ic}\Delta^{ab}_{\mu\nu}(K-Q)G_0^-(Q)\left[\Phi^+_{\bf 1}(Q)G^+_{\bf 1}(Q)\left[\mathcal{Q}_1\right]^{fg}_{cd}+\Phi^+_{\bf 8}(Q)G^+_{\bf 8}(Q)\left[\mathcal{Q}_8\right]^{fg}_{cd}\right](T_b)_{dj}\gamma^\nu
\eea
Here, we used the fact that the massless bare quark propagator is diagonal in the color-flavor space,
\bea
\left[G_0^\pm\right]^{fg}_{ij}\equiv\delta^{fg}\delta_{ij}G_0^\pm\,.
\eea
To proceed, we need to know the form of the gluon propagators including the gluon self-energies in terms of the associated spectral densities. This is presented in the next subsection.

\subsection{Gluon Propagators}\label{gluonpropagators}

In pure Coulomb gauge, the gluon propagators have the following forms:
\begin{subequations}
\bea
\Delta^{ab}_{00}(K)&\equiv&-\delta^{ab}\frac{1}{{\bf k}^2-\Pi^{cc}_{00}(K)}\,,\\
\Delta^{ab}_{0i}(K)&\equiv& 0\,,\\
\Delta^{ab}_{ij}(K)&\equiv&-\delta^{ab}\bigg(\delta_{ij}-\hat{k}_i\hat{k}_j\bigg)\frac{1}{K^2-\Pi_t^{cc}(K)}\,,
\eea
\end{subequations}
where the transverse component of the gluon self-energy is defined as
\bea
\Pi_t^{aa}(K)\equiv\frac{1}{2}\bigg(\delta^{ij}-\hat{k}^i\hat{k}^j\bigg)\Pi_{ij}^{aa}(K)\,,
\eea
In addition, the longitudinal and the transverse components of the gluon propagator are defined as $\Delta^{ab}_{00}(K)\equiv-\delta^{ab}\Delta_{00}$ and $\Delta^{ab}_{ij}(K)\equiv-\delta^{ab}\left(\delta_{ij}-\hat{k}_i\hat{k}_j\right)\Delta_t$, respectively \cite{coulomb,007}.
We insert the gluon propagators in the gap equation we have obtained so far, and use \eq{gapproj} to write the right hand side (rhs) of \eq{gap2} in terms of the singlet and the octet gaps. After some algebraic calculation and using the properties of the projectors in \eq{ppproj}, eventually, we reach the following coupled gap equations:
\begin{subequations}
\bea\label{gap1}
\Phi^+_{\bf 1}(K)&\equiv&g^2\,\frac{2T}{3V}\sum_Q\gamma^\mu\Delta^{aa}_{\mu\nu}(K-Q)G_0^-(Q)\Phi^+_{\bf 8}(Q)G_{\bf 8}^+(Q)\,\gamma^\nu\,,
\eea
\bea\label{gap8}
\Phi^+_{\bf 8}(K)&\equiv&g^2\,\frac{T}{12V}\sum_Q\gamma^\mu\Delta^{aa}_{\mu\nu}(K-Q)G_0^-(Q)\left[\Phi^+_{\bf 8}(Q)G_{\bf 8}^+(Q)+2\,\Phi^+_{\bf 1}(Q)G_{\bf 1}^+(Q)\right]\gamma^\nu
\eea
\end{subequations}
Utilizing \eq{spinor} and \eq{pron} in \eq{gap1}, we have
\bea
\Phi^+_{\bf 1}(K)&\equiv&g^2\,\frac{2T}{3V}\sum_Q\sum_{h=r,\ell}\sum_{e=\pm}\gamma^\mu\Delta^{aa}_{\mu\nu}(K-Q)\,\mathcal{P}^{-e}_{-h}\,\frac{\phi_{8,h}^e(Q)}{q_0^2-[\e_{\bf q}^e(\phi_{8,h}^e)]^2}\,\gamma^\nu
\eea
and the same procedure leads to the following expression for \eq{gap8}:
\bea\label{8}
\Phi^+_{\bf 8}(K)&\equiv&g^2\frac{T}{12V}\sum_Q\sum_{h=r,\ell}\sum_{e=\pm}\gamma^\mu\Delta^{aa}_{\mu\nu}(K-Q)\,\mathcal{P}^{-e}_{-h}\left[\frac{\phi_{1,h}^e(Q)}{q_0^2-[\e_{\bf q}^e(\phi_{1,h}^e)]^2}+\frac{2\phi_{8,h}^e(Q)}{q_0^2-[\e_{\bf q}^e(\phi_{8,h}^e)]^2}\right]\,\gamma^\nu .
\eea
Here $\mathcal{P}^{e}_{h}\equiv\mathcal{P}_h\Lambda^e_{\bf q}$. Implementation of \eq{spinor} in the left-hand side of the above equations gives
\bea\label{c1}
\phi^e_{1,h}(K)\equiv g^2\,\frac{2T}{3V}\sum_Q\Delta^{aa}_{\mu\nu}(K-Q)&\hspace{-.2cm}&\hspace{-.6cm}\Big\{\frac{\phi_{8,h}^e(Q)}{q_0^2-[\e_{\bf q}^e(\phi_{8,h}^e)]^2}\rm{Tr}\left[\mathcal{P}^e_h(k)\gamma^\mu\mathcal{P}_{-h}^{-e}(q)\gamma^\nu\right]\nonumber\\
\hspace{-.2cm}&+&\hspace{-.1cm}\frac{\phi_{8,h}^{-e}(Q)}{q_0^2-[\e_{\bf q}^{-e}(\phi_{8,h}^{-e})]^2}\rm{Tr}\left[\mathcal{P}^e_h(k)\gamma^\mu\mathcal{P}_{-h}^{e}(q)\gamma^\nu\right]\Big\}
\eea
where we used the properties of the chiral and the energy projectors. Also we find that
\bea\label{c2}
\phi^e_{8,h}(K)\equiv g^2\,\frac{T}{12V}\sum_Q\Delta^{aa}_{\mu\nu}(K-Q)&\hspace{-.2cm}&\hspace{-.6cm}\Big(\big\{\frac{\phi_{1,h}^e(Q)}{q_0^2-[\e_{\bf q}^e(\phi_{1,h}^e)]^2}+\frac{2\,\phi_{8,h}^e(Q)}{q_0^2-[\e_{\bf q}^e(\phi_{8,h}^e)]^2}\big\}\rm{Tr}\left[\mathcal{P}^e_h(k)\gamma^\mu\mathcal{P}_{-h}^{-e}(q)\gamma^\nu\right]\nonumber\\
\hspace{-.2cm}&+\hspace{-.1cm}&\big\{\frac{\phi_{1,h}^{-e}(Q)}{q_0^2-[\e_{\bf q}^{-e}(\phi_{1,h}^{-e})]^2}+\frac{2\,\phi_{8,h}^{-e}(Q)}{q_0^2-[\e_{\bf q}^{-e}(\phi_{8,h}^{-e})]^2}\big\}\rm{Tr}\left[\mathcal{P}^e_h(k)\gamma^\mu\mathcal{P}_{-h}^{e}(q)\gamma^\nu\right]\Big)
\eea
For these parts we have benefited from Ref.\cite{3} which presents the relations between the gap functions with different chiralities for the quasiparticles and the quasiantiparticles. To proceed, we apply the following identities:
\begin{subequations}
\bea
\rm{Tr}\left[\mathcal{P}_h\,\Lambda^e_{\bf k}\,\gamma_0\,\mathcal{P}_{-h}\,\Lambda^\mp_{\bf q}\gamma_0\right]\hspace{-.2cm}&\equiv\hspace{-.2cm}&\frac{1\pm\hat{{\bf k}}\cdot\hat{{\bf q}}}{2}\,,\\
\sum_i\rm{Tr}\left[\mathcal{P}_h\,\Lambda^e_{\bf k}\,\gamma_i\,\mathcal{P}_{-h}\,\Lambda^\mp_{\bf q}\gamma_i\right]\hspace{-.2cm}&\equiv\hspace{-.2cm}&-\frac{3\mp\hat{{\bf k}}\cdot\hat{{\bf q}}}{2}\\
\rm{Tr}\left[\mathcal{P}_h\,\Lambda^e_{\bf k}\,\gamma\cdot\hat{{\bf p}}\,\mathcal{P}_{-h}\,\Lambda^\mp_{\bf q}\gamma\cdot\hat{{\bf p}}\right]\hspace{-.2cm}&\equiv\hspace{-.2cm}&-\frac{1\pm\hat{{\bf k}}\cdot\hat{{\bf q}}}{2}\frac{(k\mp q)^2}{({\bf k}-{\bf q})^2}
\eea
\end{subequations}
Hence, \eq{c1} becomes
\bea\label{6}
\phi^e_{1,h}(K)\hspace{-.1cm}&\equiv\hspace{-.1cm}&g^2\frac{2T}{3V}\sum_Q\sum_{\alpha=\pm}\frac{\phi_{8,h}^{\alpha e}(Q)}{q_0^2-[\e_{\bf q}^{\alpha e}(\phi_{8,h}^{\alpha e})]^2}\Bigg\{\Delta_{00}(K-Q)\,\frac{1+\alpha\,\hat{{\bf k}}\cdot\hat{{\bf q}}}{2}\nonumber\\
&+&\Delta_t(K-Q)\,\Big[-\frac{3-\alpha\,\hat{{\bf k}}\cdot\hat{{\bf q}}}{2}+\frac{1+\alpha\,\hat{{\bf k}}\cdot\hat{{\bf q}}}{2}\frac{(k-\alpha\, q)^2}{({\bf k}-{\bf q})^2}\Big]\Bigg\}
\eea
The poles of $1/(q_0^2-[\e_{\bf q}^{\alpha e}(\phi_{8,h}^{\alpha e})]^2)$ give a residue $\sim 1/\e_{\bf q}^{\alpha e}(\phi_{8,h}^{\alpha e})$. In the weak coupling limit, in the vicinity of the Fermi surface, $k\sim \mu$, the quasiparticle energy $\e^+_{\bf q}$ is very small in comparison to the quasiantiparticle energy $\e^-_{\bf q}$ \cite{6}. Therefore, one can neglect the contribution of the quasiantiparticles in the gap equation, and, in the rhs, keep the terms with $\alpha e=+1$.
In addition, in the weak coupling limit the approximations
\begin{subequations}\label{3}
\bea
\frac{1+\hat{{\bf k}}\cdot\hat{{\bf q}}}{2}\simeq 1\,,
\eea
\bea
-\frac{3-\hat{{\bf k}}\cdot\hat{{\bf q}}}{2}+\frac{1+\hat{{\bf k}}\cdot\hat{{\bf q}}}{2}\,\frac{(k\mp q)^2}{({\bf k}-{\bf q})^2}\simeq -1\,.
\eea
\end{subequations}
are accepted. We now add and subtract the HDL gluon propagators to \eq{6}:
\bea\label{4}
\phi^e_{1,h}(K)\hspace{-.2cm}&\equiv\hspace{-.2cm}&g^2\frac{2T}{3V}\sum_Q\sum_{\alpha=\pm}\frac{\phi_{8,h}^{\alpha e}(Q)}{q_0^2-[\e_{\bf q}^{\alpha e}(\phi_{8,h}^{\alpha e})]^2}\bigg(\Delta^{{\rm HDL}}_{00}(K-Q)-\Delta^{{\rm HDL}}_t(K-Q)\bigg)+\delta\phi^e_{1,h}(K)\,,
\eea
where $\delta\phi^e_{1,h}(K)$ is the difference between the gap equation in the color-superconducting phase with that in the HDL limit. Hence, for $\delta\phi^e_{1,h}(K)=0$ we recover the standard gap equation with the gluon propagators in HDL approximation. In the weak coupling limit, the HDL solution up to subleading order is presented in Ref.\,\cite{7}. The extra parts in \eq{4} are the correction to the previous result.

To take the gluon self-energy effects into account, we have to write the gluon propagators in terms of their spectral densities. This is done in the following.

\subsection{Spectral Densities}\label{spectraldensities}

We now need to go to the spectral representation of the gluon propagators
\begin{subequations}\label{2}
\bea
\Delta_{00}(K)&\equiv&-\frac{1}{k^2}+\int_0^{1/T}\,d\tau\, {\rm{e}}^{ik_0\tau}\Delta_{00}(\tau,{\bf k})\,,\\
\Delta_t(K)&\equiv&\int_0^{1/T}\,d\tau\, {\rm{e}}^{ik_0\tau}\Delta_t(\tau,{\bf k})\,,\\
\Delta_{00,t}(\tau,{\bf k})&\equiv&\int_0^\infty d\omega\,\rho_{00,t}(\omega,{\bf k})\left\{\left[1+n_B(\omega/T)\right]{\rm{e}}^{-\omega\tau}+n_b(\omega/T){\rm{e}}^{\omega\tau}\right\}\,,
\eea
\end{subequations}
where $n_B(x)\equiv 1/({\rm{e}}^x-1)$ is the Bose-Einstein distribution function. The term $-1/k^2$ in the electric propagator cancels the contribution of $\Delta_{00}(K)$ when $k_0\rightarrow \infty$. This term is the same for the electric gluon propagator in a superconductor and the electric HDL propagator, because there are not any differences between these propagators for $k_0\ll \phi$ \cite{spectralrep,10}. The spectral densities are defined as
\bea
\rho_{00,t}(p_0,{\bf k})\equiv\frac{1}{\pi}{\rm Im}\,\Delta_{00,t}(p_0+i\eta,{\bf k})\,.
\eea
When ${\rm Im}{\hat \Pi}_{00}(p_0,{\bf p})$ and ${\rm Im}\Pi_{t}(p_0,{\bf p})$ are nonzero, the spectral densities are regular and the above equation is identical to
\bea\label{spectral1}
\rho_{00}(p_0,{\bf p})\hspace{-.2cm}&\equiv\hspace{-.2cm}&\frac{1}{\pi}\frac{{\rm Im}{\hat \Pi}_{00}(p_0,{\bf p})}{\left[{\bf p}^2-{\rm Re}\,{\hat \Pi}_{00}(p_0,{\bf p})\right]^2+\left[{\rm Im}\,{\hat \Pi}_{00}(p_0,{\bf p})\right]^2}\,,\nonumber\\
\rho_{t}(p_0,{\bf p})\hspace{-.2cm}&\equiv\hspace{-.2cm}&\frac{1}{\pi}\frac{{\rm Im}\Pi_{t}(p_0,{\bf p})}{\left[p_0^2-{\bf p}^2-{\rm Re}\,\Pi_{t}(p_0,{\bf p})\right]^2+\left[{\rm Im}\,\Pi_{t}(p_0,{\bf p})\right]^2}\,,
\eea
where ${\hat \Pi}_{00}$ is the modified longitudinal gluon self-energy in the CFL phase, cf. Ref.\,\cite{mal1}. When ${\rm Im}{\hat \Pi}_{00}(p_0,{\bf p})={\rm Im}\Pi_{t}(p_0,{\bf p})= 0$, for a given momentum ${\bf p}$, the spectral densities have poles determined by
\begin{subequations}
\bea
\left[{\bf p}^2-{\rm Re}\,{\hat \Pi}_{00}(p_0,{\bf p})\right]_{p_0=\omega_{00}({\bf p})}=0\,,\\
\eea
for the electric gluons, and
\bea
\left[p_0^2-{\bf p}^2-{\rm Re}\,\Pi_{t}(p_0,{\bf p})\right]_{p_0=\omega_{t}({\bf p})}=0\,,
\eea
\end{subequations}
for the magnetic gluons. The quasiparticle dispersion relation can be obtained by the solutions of $p_0= \omega_{00,t}({\bf p})$. Therefore in this case
\bea\label{spectral2}
\rho_{00,t}(p_0,{\bf p})&\equiv&-\mathcal{Z}_{00,t}({\bf p})\big\{\delta\left[\,p_0-\omega_{00,t}({\bf p})\,\right]-\delta\left[\,p_0+\omega_{00,t}({\bf p})\,\right]\big\}\,,
\eea
where $\mathcal{Z}_{00,t}({\bf p})$ are the residues at the poles $p_0=\omega_{00,t}({\bf p})$,
\bea\label{poles}
\mathcal{Z}_{00,t}({\bf p})\equiv\left(\left|\frac{\partial(\Delta_{00,t})^{-1}(p_0,{\bf p})}{\partial p_0}\right|\right)^{-1}_{p_0=\omega_{00,t}({\bf p})}\,.
\eea

Now we can go back to \eq{4} and continue the calculations. We can write
\bea\label{1}
\frac{\phi_i(Q)}{q_0^2-\e_{\bf q}^2(\phi_i)}&=&-\int_0^{1/T} d\tau\, {\rm{e}}^{q_0\tau}\frac{\phi_i(\e_{\bf q},{\bf q})}{2\e_{\bf q}(\phi_i)}\left(\left\{1-n_F\left[\e_{\bf q}(\phi_i)/T\right]\right\}{\rm{e}}^{-\e_{\bf q}(\phi_i)\tau}-n_F\left[\e_{\bf q}(\phi_i)/T\right]{\rm{e}}^{\e_{\bf q}(\phi_i)\tau}\right)\,,
\eea
where $n_F(x)\equiv 1/({\rm e}^x+1)$ is the Fermi-Dirac distribution function. Now, the Matsubara sum should be performed over the fermionic energies $q_0=-i(2n+1)\pi T$. Neglecting the quasiantiparticle contributions and inserting Eqs.\,(\ref{2}) and (\ref{1}) in \eq{4} we have
\bea\label{5}
\delta\phi_{k,1}\hspace{-.15cm}&\equiv\hspace{-.15cm}& -\frac{2}{3}g^2\int_0^\infty d\omega\int \frac{d^3 q}{(2\pi)^3}\left(\delta\rho_{00}-\delta\rho_t\right)\frac{\phi_{q,8}}{2\,\e_{{\bf k},8}}\hspace{-.1cm}\left(\frac{1}{\omega+\e_{{\bf q},8}+\e_{{\bf k},8}}+\frac{1}{\omega+\e_{{\bf q},8}-\e_{{\bf k},8}}\right)
\eea
where, for simplicity, we dropped the subscript $h$ and the superscript $+$, abbreviated $\phi_i(\e_{\bf q},{\bf q})\equiv\phi_{q,i}$, and set $\e_{\bf q}(\phi_{q,i})=\e_{{\bf q},i}$. We have also made an analytical continuation onto the quasiparticle mass shell, $q_0\rightarrow\e_{\bf q}+i\eta$. In \eq{5}
\bea\label{10}
\delta\rho_{00,t}\equiv\rho_{00,t}-\rho_{00,t}^{{\rm HDL}}\,.
\eea
Making use of the fact that the spectral densities are isotropic, we can change the integral variable ${\rm cos}\theta\equiv{\hat{\bf k}}\cdot{\hat{\bf q}}$ to $p$. Then, \eq{5} reads
\bea\label{7}
\delta\phi_{k,1}\hspace{-.15cm}&\simeq\hspace{-.15cm}&\frac{g^2}{12}\int_{\mu-\delta}^{\mu+\delta}dq\,
\int_0^\infty d\omega\left[\,\mathcal{I}_{t}(\omega)-\mathcal{I}_{00}(\omega)\,\right]\frac{\phi_{q,8}}{\e_{{\bf q},8}}\left(\frac{1}{\omega+\e_{{\bf q},8}+\e_{{\bf k},8}}+\frac{1}{\omega+\e_{{\bf q},8}-\e_{{\bf k},8}}\right)
\eea
where
\bea\label{t}
\mathcal{I}_{00,t}(\omega)\equiv\int_0^{2\mu}dp\,\,p\,\delta\rho_{00,t}(\omega,p)\,.
\eea
To derive these formulas, we have employed the approximations $k/q\simeq 1$, $|k-q|\simeq 0$, and $k+q\simeq 2\mu$. These approximations are justified because the $q$-integral peaks at the Fermi surface \cite{rischke094}. Moreover, since the gap is strongly dependent on $q$, the momentum integration is restricted to the Fermi surface $\mu-\delta\le q\le\mu+\delta$.

Note that the functions $\mathcal{I}_{00,t}(\omega)$ are dimensionless. Therefore, all quantities derived from them must be dimensionless as well, e.g. $\omega/m_g$, $\phi/m_g$, $\omega/\mu$, $\phi/\mu$, etc. We will make use of this property of $\mathcal{I}_{00,t}(\omega)$ in the coming sections.

The same procedure for \eq{8} yields
\bea\label{20}
\delta\phi_{k,8}\simeq\frac{g^2}{96}\int_{\mu-\delta}^{\mu+\delta}dq\,
\int_0^\infty d\omega\left[\,\mathcal{I}_{t}(\omega)-\mathcal{I}_{00}(\omega)\,\right]\hspace{-.15cm}&&\hspace{-.4cm}\left[\frac{\phi_{q,1}}{\e_{{\bf q},1}}\left(\frac{1}{\omega+\e_{{\bf q},1}+\e_{{\bf k},1}}+\frac{1}{\omega+\e_{{\bf q},1}-\e_{{\bf k},1}}\right)\right.\nonumber\\
&+&\left.2\frac{\phi_{q,8}}{\e_{{\bf q},8}}\left(\frac{1}{\omega+\e_{{\bf q},8}+\e_{{\bf k},8}}+\frac{1}{\omega+\e_{{\bf q},8}-\e_{{\bf k},8}}\right)\right]\nonumber\\
\eea
From now on we apply the approximation $\phi_{1,q}\approx 2\,\phi_{8,q}\approx 2\phi_q$ (cf. Ref.\cite{gap}) in the previous expressions. Doing so, for \eq{7} we have
\bea\label{21}
\delta\phi_{k}\hspace{-.15cm}&\simeq\hspace{-.15cm}&\frac{g^2}{24}\int_{\mu-\delta}^{\mu+\delta}dq\,
\int_0^\infty d\omega\left[\,\mathcal{I}_{t}(\omega)-\mathcal{I}_{00}(\omega)\,\right]\frac{\phi_{q}}{\e_{{\bf q}}}\left(\frac{1}{\omega+\e_{{\bf q}}+\e_{{\bf k}}}+\frac{1}{\omega+\e_{{\bf q}}-\e_{{\bf k}}}\right)\,,
\eea
where we set $\e_{{\bf q},8}=\e_q$. In the rest of this paper we study the gluon self-energy effects on the value of only the octet gap function [\eq{21}]. All the arguments presented in the next sections are valid for the singlet gap function as well [\eq{20}].

Now, we have to know the values of the HDL and the CFL spectral densities for different values of the energy and the momentum. We discuss this in great detail in Sec. \ref{dispersionrelations}.

\subsection{Dispersion Relations}\label{dispersionrelations}

\begin{figure}
\includegraphics[width=10cm]{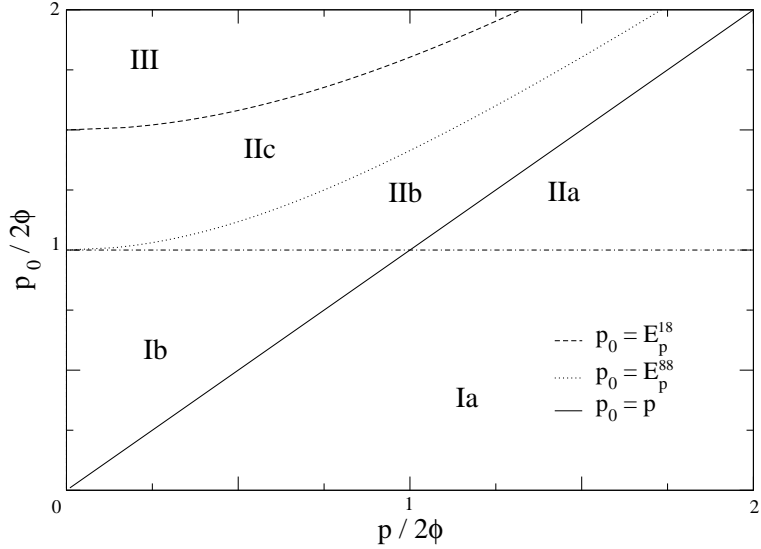}
\caption
{Schematic plot of the $(p_0,p)$ plane. In regions Ia and Ib, the HDL spectral densities are calculated from \eq{spectral1}, and those in the rest of the plane, from \eq{spectral2}. The CFL spectral densities are give by \eq{spectral1} in regions IIa, IIb, IIc, and III. In regions Ia and Ib, where the light plasmon mode exists, \eq{spectral2} is used to evaluate the spectral density. ${\rm E}^{18}_p=\sqrt{p^2+9\phi_p^2}$ and ${\rm E}^{88}_p=\sqrt{p^2+4\phi_p^2}$, cf. \cite{mal1}.
}
\label{fig}
\end{figure}

Below the light cone, $p_0 < p$, the imaginary part of the HDL electric and magnetic gluons is nonzero, cf. Fig.(1) of Ref.\,\cite{mal1}. Hence, the gluons are Landau damped. This is equivalent to regions Ia and IIa of Fig.\,(\ref{fig}) of this paper. In these regions, then, the spectral densities are calculated from \eq{spectral1}. However, above the light cone, $p_0 > p$, regions IIb, IIc, and III, the imaginary parts are zero, and therefore, one has to find the spectral densities using \eq{spectral2}. The dispersion relation of the electric and the magnetic gluons in this case are obtained from $p_0= \omega_{00}({\bf p})$ and $p_0=\omega_t({\bf p})$, respectively. The weak coupling limit, $m_g\ll \phi$, corresponds to region III of Fig.\,(\ref{fig}), where $\omega_{00}(0)=\omega_t(0)=m_g$.

In the CFL phase, below the light cone and for $p_0<2\phi$, region Ia of Fig.\,(\ref{fig}), the imaginary part of the electric and magnetic gluons are zero; one uses \eq{spectral2} to find the spectral densities. However, below the light cone and for $p_0>2\phi$, the imaginary parts are not zero. This is the case for $p_0>p$ as well. In both of these cases, the spectral densities are obtained from \eq{spectral1}, cf. Ref.\,\cite{mal1}. Although the imaginary parts in regions IIb, $p<p_0<E_{\bf p}^{88}$, and IIc, $E_{\bf p}^{88}<p_0<E_{\bf p}^{18}$, are approximately at order of the HDL limit, they are very small in region III, $p_0>E_{\bf p}^{18}$. In this region ${\rm Im}{\hat \Pi}_{00}(p_0,{\bf p})\sim{\rm Im}\Pi_{t}(p_0,{\bf p})\sim\phi^2$.

The Nambu-Goldstone (NG) bosons appear in region Ia, cf. Ref.\,\cite{mal1}. The light plasmon modes, however, are present in regions Ia and Ib. In these regions, as stated before, the imaginary parts of the color-superconducting gluon self-energy are zero and the gluon energy and momentum are much smaller than the gap. The dispersion relation of the NG bosons is evaluated from $P_\mu P_\nu\Pi^{\mu\nu}(P)=0$ and that for the light plasmons is calculated from \eq{spectral2}. In this regime, $p_0, p\ll\phi$, the dispersion relation is approximately linear. Hence, we can expand the self-energies in terms of its momentum and energy. The leading terms are
\begin{subequations}\label{selfapp}
\bea
\Pi^{00}(p_0,{\bf p})&\simeq& \frac{m_g^2}{3}\left(2.6+0.4\,\frac{p_0^2}{\phi^2}\right)\,,\\
\Pi^{\ell}(p_0,{\bf p})&\simeq& \frac{m_g^2}{10}\,\frac{p_0^2}{\phi^2}\,,\\
\Pi^{0i}(p_0,{\bf p}){\hat p}^i&\simeq& -\frac{m_g^2}{150}\,\frac{p}{\phi}\,\frac{p_0^2}{\phi^2}\,,\\
\Pi^{t}(p_0,{\bf p})&\simeq& \frac{m_g^2}{20}\left(1-3\,\frac{p_0^2}{\phi^2}\right)\,,
\eea
\end{subequations}
where the explicit forms of the self-energies are taken from Eq.\,(A2) of Ref.\,\cite{mal1}. We refrain from presenting the details of the calculations leading to Eq.\,(\ref{selfapp}). Similar calculations for the 2SC phase are done in Ref.\,\cite{sonrischke}. In this limit, the longitudinal self-energy ${\hat \Pi}^{00}(p_0,{\bf p})$ [cf. Eq.(4b) of Ref.\,\cite{mal1}] is given by
\bea
{\hat \Pi}^{00}(p_0,{\bf p})\simeq \frac{m_g^2}{10}\,\frac{p^2}{\phi^2}\left(1+0.2\,\frac{p_0^2}{\phi^2}\right)\,.
\eea
Having calculated the transverse self-energy in Eq.\,(\ref{selfapp}), the magnetic gluon propagator, $\Delta^{t}(P)=1/(p_0^2-p^2-\Pi^{t}(P))$ becomes
\bea\label{poll}
\left[\Delta^{t}(P)\right]^{-1}\simeq\left[p_0^2\left(1+\frac{3}{20}\frac{m_g^2}{\phi^2}\right)-{\bf p}^2-\frac{m_g^2}{20}\right]\,.
\eea
The light plasmon dispersion relation, therefore, reads
\bea\label{omegat}
\omega_t\simeq\sqrt{\frac{20}{3}}\frac{\phi}{m_g}\sqrt{p^2+\frac{m_g^2}{20}}\,.
\eea
Employing this and \eq{poll} in \eq{poles} one can find the associated residue
\bea\label{z}
\mathcal{Z}_t({\bf p})\simeq\sqrt{\frac{5}{3}}\frac{\phi}{m_g p}\,.
\eea
We will make use of this result in Sec. \ref{sphi}.

In the next section we estimate the effect of the gluon self-energy on the value of the gap. For that, we have to investigate the spectral densities in the HDL limit as well as in the color-superconducting phase, cf. Equation (\ref{t}).

\section{Results}\label{qualitativeresults}

In this section, I estimate the values of $\mathcal{I}_{00}(\omega)$ and $\mathcal{I}_{t}(\omega)$ to find the effects of the self-energy on the value of the gap.

\subsection{Estimate for $\mathcal{I}_{00}(\omega)$}\label{estimate00}

The qualitative results of this part are very similar to those of the 2SC phase for $\mathcal{I}_{00}^{11}(\omega)$ in Ref.\,\cite{rischke094}, except that region IIc in Fig.\,(\ref{fig}) is new for the CFL phase. For $p_0<2\phi$, the spectral densities of the CFL phase are zero for both regions Ia and Ib, $\rho_{00}=0$, cf. Fig.(3) of \cite{mal1}. For the HDL limit, above the light cone, in region Ib, the spectral density is zero, therefore $\delta\rho_{00}=0$. However, in region Ia, the HDL spectral density is regular and we have $\delta\rho_{00}=-\rho_{00}^{\rm HDL}$. In this case, the spectral density is given by
\bea\label{222}
\rho_{00}^{\rm HDL}(\omega,p)=\theta(p-\omega)m_g^2\frac{\omega}{p}\left\{\left[p^2+2\,m_g^2\left(1-\frac{\omega}{2p}{\rm ln}\left|\frac{p+\omega}{\p-\omega}\right|\right)\right]^2+\left(\pi m_g^2\frac{\omega}{p}\right)^2\right\}^{-1}\,,
\eea
where $m_g$ is the gluon mass at zero temperature, $m_g^2=g^2\mu^2/2\pi^2$. For $p$ at order of $\omega$, the HDL spectral density, \eq{222}, is approximately zero and for $p\gg \omega$ it is of order $\sim 1/p^5$. Therefore, using \eq{t} we have
\bea
\mathcal{I}_{00}(\omega)\sim\frac{\omega}{m_g}\sim\frac{\phi}{m_g}\,.
\eea
Here, we neglected powers of $g$ because they do not have any consequence on the order of $\mathcal{I}_{00}(\omega)$, cf.\cite{rischke094}. The effect of color superconductivity, however, appears for $p_0>2\phi$. In region IIa, both $\rho_{00}$ and $\rho^{\rm HDL}_{00}$ are regular. Since for $\phi\rightarrow 0$ we recover the normal phase, $\delta\rho_{00}\rightarrow 0$ (cf. Ref.\,\cite{mal1}), to leading we have
\bea
\int_\omega^{2\mu}dp\,p\,\delta\rho_{00}(\omega,p)\sim\frac{\phi}{m_g}\,.
\eea
On the other hand, in regions IIb and IIc, the HDL spectral densities are zero and $\rho_{00}$ is of order $1/m_g^2$. Then, for region IIb
\bea
\int_{\sqrt{\omega^2-4\,\phi^2}}^\omega dp\,p\,\delta\rho_{00}(\omega,p)\sim\frac{\phi^2}{m_g^2}\,,
\eea
and for region IIc
\bea
\int_{\sqrt{\omega^2-9\,\phi^2}}^{\sqrt{\omega^2-4\,\phi^2}} dp\,p\,\delta\rho_{00}(\omega,p)\sim\frac{\phi^2}{m_g^2}\,.
\eea
In region III, the spectral density for the CFL phase is a smeared delta function and that for the HDL limit is a true one. Nevertheless, the integral over momentum in \eq{t} makes $\mathcal{I}_{00}(\omega)$ regular. Qualitatively, the argument used for region IIa is valid at this region too. When $\phi\rightarrow 0$ we have $\rho_{00}\simeq\rho_{00}^{\rm HDL}$, hence,
\bea
\int_0^{\sqrt{\omega^2-9\,\phi^2}} dp\,p\,\delta\rho_{00}(\omega,p)\sim\frac{\phi}{m_g}\,.
\eea
In conclusion, $\mathcal{I}_{00}(\omega)$ is at most of order $\phi/m_g$.

\subsection{Estimate for $\mathcal{I}_{t}(\omega)$}\label{estimatet}

The arguments used in the previous section for regions where $p_0>2\phi$ are valid for $\mathcal{I}_{t}(\omega)$ too. However, for $p_0<2\phi$, the scenario is different because the light plasmon modes appear in the regions Ia and Ib. In these regions, since the imaginary parts of the gluon self-energy are zero, one has to use \eq{spectral2} to find the associated spectral densities. For both of these regions, the color-superconducting contribution is
\bea\label{intt}
\int_\omega^{2\mu}dp\, p\, \rho_t(\omega,p)=-p(\omega_t)\mathcal{Z}_t(p)\left(\left|\frac{\partial\,
\omega_t(p)}{\partial\,p}\right|\right)^{-1}_{p=p(\omega_t)}
\eea
Making use of \eqq{omegat}{z} in \eq{intt} yields
\bea
\int_\omega^{2\mu}dp\, p\, \rho_t(\omega,p)\simeq -1\,.
\eea
In region Ia, one can approximate the HDL spectral density for $p\gg\omega$ as
\bea
\rho_t^{\rm HDL}(\omega,p)\simeq\theta(p-\omega)\,m_g^2\,\frac{\omega\, p}{p^6+(\pi\, m_g^2\, \omega)^2}\,.
\eea
This gives rise to
\bea
\int_\omega^{2\mu}dp\, p\, \rho_t^{\rm HDL}(\omega,p)\simeq\frac{\phi}{m_g}\,.
\eea
Therefore in this region, to leading order, $\mathcal{I}_{t}(\omega)$ is at most of order one,
\bea\label{15}
\mathcal{I}_{t}(\omega)=\int_\omega^{2\mu}dp\, p\, \delta\rho_t(\omega,p)\sim -1\,.
\eea
Furthermore, since in region Ib the HDL spectral density vanishes whereas the light plasmon spectral density persist, $\mathcal{I}_{t}(\omega)$ is of order $\sim 1$ too.

As mentioned, for the other regions, $p_0>2\phi$, $\mathcal{I}_{t}(\omega)$ has the same order of magnitude as $\mathcal{I}_{00}(\omega)$, namely, it is at most of order $\sim\phi/m_g$. In the next section, we take into account these effects on the value of the gap.

\subsection{Estimate for $\delta\phi_k$}\label{sphi}

Having estimated the order of $\mathcal{I}_{00}(\omega)$ and $\mathcal{I}_{t}(\omega)$ we can now estimate the effect of the gluon self-energy on the gap. For $p_0<2\phi$, $\mathcal{I}_{00}(\omega)$ is at most of order $\sim\phi/m_g$ and $\mathcal{I}_{t}(\omega)$ of order $\sim 1$. As mentioned earlier, we do not take into account the additional powers of the coupling $g$, because they are always accompanied by at least one power of the gap which is exponentially small in $g$, $\phi\sim\exp(-1/g)$. Employing \eq{15} in \eq{21}, after performing the $\omega$ integral, we have
\bea
\delta\phi_{k}\sim -g^2\int_{\mu-\delta}^{\mu+\delta}dq\,\frac{\phi_{q}}{\e_{{\bf q}}}\,{\rm ln}\left|\frac{(2\phi+\e_{{\bf q}})^2-\e_{{\bf k}}^2}{\e_{{\bf q}}^2-\e_{{\bf k}}^2}\right|\,,
\eea
In Appendix B of Ref.\cite{rischke094} it is shown that this integral has a positive value and is at most of order $\sim \phi$. Therefore, $\delta\phi_k$ contributes to the $O(g)$ of the gap in \eq{qcdgap}, i.e. to the subsubleading order. Thus, the coefficients $b$ and $c$ are not modified. In addition, since $\delta\phi_k$ has a negative sign, the effect of the gluon self-energy is to decrease the value of the gap. This is in agreement with the previous result obtained using the Ginzburg-Landau theory, which shows near the critical temperature $T_c$ the CFL gluon self-energy decreases the value of the gap \cite{iida}. 

\section{Conclusion}

In conclusion, in the weak coupling limit, we studied the effect of CFL gluon self-energy on the solution of the gap. For the values of the energy and momentum, where the spectral densities of the the CFL color superconductivity are regular, the effect of the gluon self-energy is very negligible, i.e. it does not even appear up to subsubleading order. However, for the energies and momenta, where the light plasmon mode appears, the effect is of subsubleading order.

\section{Acknowledgments}

The author thanks Owe Philipsen for fruitful discussions and also Institute for Theoretische Physik of Muenster University for financial support.

\end{document}